\documentclass[aps,prx,twocolumn,superscriptaddress,numerical,showpacs]{revtex4-1}

\usepackage{tensor}
\usepackage{slashed}
\usepackage{epsfig}
\usepackage{amsmath}
\usepackage{amssymb}
\usepackage{graphicx}% Include figure files
\usepackage{dcolumn}% Align table columns on decimal point
\usepackage{bm}% bold math
\usepackage{bbold}
\usepackage{hyperref}% add hypertext capabilities
\hypersetup{
     colorlinks   = true,
     citecolor    = blue,
     urlcolor = blue,
     linkcolor = blue
}

\begin{document}

\title{Mixed axial-torsional anomaly in  Weyl semimetals}% Force line breaks with \\

\author{Yago Ferreiros}
  \affiliation{Department of Physics, KTH Royal Institute of Technology, 106 91 Stockholm, Sweden}%Lines break automatically or can be forced with \\
  \email{ferreiros@kth.se}
\author{Yaron Kedem}
  \affiliation{Department of Physics, Stockholm University, AlbaNova University Center, 106 91 Stockholm, Sweden}
\author{Emil J. Bergholtz}
  \affiliation{Department of Physics, Stockholm University, AlbaNova University Center, 106 91 Stockholm, Sweden}
\author{Jens H. Bardarson}
  \affiliation{Department of Physics, KTH Royal Institute of Technology, 106 91 Stockholm, Sweden}
\date{\today}% It is always \today, today,

\begin{abstract}
We show that Weyl semimetals exhibit a mixed axial-torsional anomaly in the presence of axial torsion, a concept exclusive of these materials with no known natural fundamental interpretation in terms of the geometry of spacetime.
This anomaly implies a nonconservation of the axial current---the difference in current of left- and right-handed chiral fermions---when the torsion of the spacetime in which the Weyl fermions move couples with opposite sign to different chiralities.
The anomaly is activated by driving transverse sound waves through a Weyl semimetal with a spatially varying tilted dispersion, which can be engineered by applying strain. 
This leads to sizable alternating current in presence of a magnetic field that provides a clear-cut experimental signature of our predictions.  
\end{abstract}

\maketitle

%\tableofcontents

{\it Introduction.}---The recently realized Weyl semimetals~\cite{WTVS11,WFBD15,HXBH15,H115,LXD15,LW15,SNY15,YLC15,XAH15,GB16} are gapless three-dimensional topological materials whose low energy excitations are Weyl fermions.
% . 
In field theory, Weyl fermions exhibit the chiral anomaly~\cite{A69,BJ69}, the phenomena that their current in the presence of nonorthogonal electric $\vec{E}$ and magnetic $\vec{B}$ fields is not conserved at the quantum level.
The continuity equation for the four-current $J^\mu_{L/R}$ of a single Weyl fermion of a given chirality (left or right handed) reads~\cite{ZYZ84} $\partial_\mu J^\mu_{L/R}=\pm\, e\,\vec{E}\cdot\vec{B}/12\pi^2\hbar^2$.
In the condensed matter realization of Weyl semimetals, the Weyl fermions necessarily come in pairs of opposite chirality~\cite{NN281}.
Due to the chiral anomaly, their associated currents are not individually conserved; the freedom of having independent gauge fields coupling to each chirality means that a cancellation of the chiral anomaly between the two chiralities does not necessarily happen, and conservation of the total vector current $J^\mu = J_L^\mu + J_R^\mu$ is not guaranteed \cite{L16}. 
Vector current conservation can be recovered by picking a specific, left-right asymmetric regularization of the underlying quantum field theory, which as a consequence results in the axial current $J_5^\mu = J_L^\mu - J_R^\mu$ not being conserved~\cite{A69,BJ69,ZB12,L14,BKY14,L16}: $\partial_\mu J^\mu_{5}=e\,\vec{E}\cdot\vec{B}/2\pi^2\hbar^2$.
This nonconservation of the axial current is referred to as the axial anomaly (in the condensed matter literature it is often also called the chiral anomaly). 
It is important to note that at the field theory level the chiral anomaly only forbids the simultaneous conservation of the vector and axial currents, but it is a natural physical choice to impose conservation of the vector current.  
The axial anomaly is predicted to result in a negative magnetoresistance in Weyl semimetals~\cite{NN83,SS13,HQ13,GMS14}, which  has been experimentally observed~\cite{ZXHJ16,XCO15,HZDC15,LCO18}.

The aforementioned freedom to have independent gauge fields for each chirality means that the axial anomaly gets a contribution beyond the electromagnetic one.
This contribution occurs in the presence of axial gauge fields, $A^5_\mu = (A^{L}_\mu - A^{R}_\mu)/2$, which arise in Helium-3~\cite{Bevan1997,KV05} and are induced for example by strain or inhomogeneous magnetization~\cite{CFLV15,LYQ13} in Weyl semimetals.
The axial fields couple to the two chiralities with opposite sign, and in analogy to electromagnetic fields give rise to a continuity equation for the axial current of the form~\cite{L16,BRKBG18} $\partial_\mu J^\mu_{5}=e\,\vec{E}^5\cdot\vec{B}^5/6\pi^2\hbar^2$, where the axial electric and magnetic fields, $\vec{E}^5$ and $\vec{B}^5$, are obtained from $A_\mu^5$ analogously to their electromagnetic counterparts.
This leads to alternative signatures for the axial anomaly in strained Weyl semimetals \cite{PCF16,GVVI16}, while the mere presence of the axial gauge fields implies a plethora of new phenomena in Weyl semimetals~\cite{CKLV16,CZ16,KZ16,LPF17,GMSS17,GMSS217,LFF17,PI18,GM18}.

A further, less studied, contribution to the axial anomaly results from torsion of spacetime.
An intuitive notion of torsion comes from its effect on vector fields: vectors are twisted when parallel transported around a curve in a differential manifold with torsion~\cite{HHKN76}.
While there is no experimental evidence for torsion in the spacetime of our universe, extensions of general relativity that include torsion, such as the  Einstein-Cartan theory~\cite{C22}, exist.
In condensed matter, torsion is, however, allowed and has been discussed in the context of Weyl semimetals~\cite{PHL14,LS14,Z15,YCH16,Y16,SF16,CZ17,HLZZ18}, topological insulators~\cite{HLF11,HLP13,PHL14}, graphene~\cite{JCV10}, and Helium-3~\cite{IMTF18}.
Since torsion affects spacetime, it influences the energy-momentum tensor, which for a single Weyl fermion cannot be jointly conserved with the electric current. This obstruction to a simultaneous conservation of energy-momentum and current is usually referred to as a mixed anomaly.
However, as before, in the presence of pairs of Weyl fermions of opposite chirality one can impose conservation of both energy-momentum and current, at the cost of nonconservation of the axial current, which now acquires torsional corrections beyond the (axial) electromagnetic contributions \cite{CZ97,PHL14,HLZZ18}---this is the mixed axial-torsional anomaly.
In addition, spacetime curvature can result in gravitational contributions to the axial anomaly~\cite{GW84}---we do not discuss these here.

A natural question now arises: is there, analogously to axial electromagnetic fields, a notion of axial torsional fields, and if so, do they give rise to new additional terms in the axial anomaly?
At the face of it, the answer would seem to be no, since torsion is a property of spacetime and as such does not know about chiralities, at least not at a fundamental level.
However, we show in this work that in a material such as a Weyl semimetal, axial torsion is realized under the application of strain. 
We derive the resulting mixed axial-torsional anomaly with axial torsion, and propose a realistic experimental setup that activates it. This work constitutes the first proposal for the realization and measurement of torsional contributions to the axial anomaly.

{\it Mixed axial-torsional anomaly.---}
In a system consisting of a pair of left- and right-handed Weyl fermions, the torsional contribution to the axial anomaly reads
\begin{equation}
\partial_\mu J_5^\mu=\frac{e}{16\pi^2l^2}\epsilon^{\mu\nu\rho\lambda}\,\Big(T_{\mu\nu}^aT_{\rho\lambda}^b+\frac{l^2}{l^2_5}\,T_{\mu\nu}^{5,a}T_{\rho\lambda}^{5,b}\Big)\,\eta_{ab},
\label{eq:axial-torsional-anomaly}
\end{equation}
where $\eta_{ab}$ is the Minkowsky metric, $a,\mu=t,x,y,z$, and $T_{\mu\nu}^{a}$ and $T_{\mu\nu}^{5,a}$ are the torsion and axial torsion tensor respectively, which we define below. 
The first term in Eq.~\eqref{eq:axial-torsional-anomaly} is known as the Nieh-Yan term~\cite{NY82,CZ97}, while the second axial torsion term is new and is our main result. 
The derivation of the axial torsion term proceeds similarly to that of the axial anomaly in the presence of axial gauge fields~\cite{L16}:
We start from the known expression~\cite{CZ97} of the mixed axial-torsional anomaly of Weyl fermions, and allow for axial torsion.
This directly results in an apparent nonconservation of both axial and vector currents; to restore vector-current conservation we need to depart from a left-right symmetric regularization by introducing Bardeen counterterms \cite{B69}, resulting in Eq.~\eqref{eq:axial-torsional-anomaly}.
For details we refer the reader to appendix \ref{appendix}.
An important difference to the universal electromagnetic contributions to the axial anomaly, is that the torsional contributions are nonuniversal and depend explicitly on the regularization through the cut-off length scales $l$ and $l_5$. 
Moreover, different regularizations, still respecting current and energy-momentum conservation, characterized by additional Bardeen counterterms, change the coefficient of the axial torsion term, such that even the ratio $l/l_5$ is nonuniversal. 

To define the (axial) torsion tensor, we introduce a set of four orthonormal basis vectors $\underline{e}_a^\mu$, one for each spacetime component $a$, at each point of the manifold (see for example \cite{HHKN76,C97}). 
Being an orthonormal basis the vectors fulfill $g_{\mu\nu}\underline{e}_a^\mu\underline{e}_b^\nu=\eta_{ab}$, where $g_{\mu\nu}$ is the (covariant) metric tensor. 
$\underline{e}_a^\mu$ is usually referred to as the frame field, and we define its inverse, the coframe field $e^a_\mu$, such that $\underline{e}_a^\mu e^a_\nu=\delta^\mu_\nu$. 
In terms of these fields, the contravariant and covariant metrics are $g^{\mu\nu}=\underline{e}_a^\mu\underline{e}_b^\nu\eta^{ab}$ and $g_{\mu\nu}=e^a_\mu e^b_\nu\eta_{ab}$. 
The torsion tensor is simply defined as the field strength of the coframe field $T^a_{\mu\nu}=\partial_\mu e^a_\nu-\partial_\nu e^a_\mu$; in analogy with the electromagnetic field we can then define a set of four (one for each spacetime component $a$) torsional electric and magnetic fields $\mathcal{E}_i^a=\partial_te^a_i-\partial_ie^a_t$ and $\mathcal{B}_i^a=\epsilon^{ijk}\partial_je^a_k$, where $i=x,y,z$. 
Similarly, we define the axial torsion tensor to be the field strength of the axial coframe field $e_\mu^{5,a}=(e_\mu^{L,a}-e_\mu^{R,a})/2$, where we allow for the possibility that left- and right-handed fermions have different coupling to the background geometry, described by the two distinct frame fields $e_\mu^{L/R,a}$.
The axial torsional electric and magnetic fields are then given by $\mathcal{E}_i^{5,a}=\partial_te^{5,a}_i-\partial_ie^{5,a}_t$ and	 $\mathcal{B}_i^{5,a}=\epsilon^{ijk}\partial_je^{5,a}_k$, and the anomaly Eq.~\eqref{eq:axial-torsional-anomaly} can be written as $2\pi^2l^2\partial_\mu J_5^\mu=e\,\vec{\mathcal{E}}_a\cdot\vec{\mathcal{B}}^{\,a}+el^2/l_5^2\,\vec{\mathcal{E}}_a^{\,5}\cdot\vec{\mathcal{B}}^{\,5,a}$.
Activating the mixed axial-torsional anomaly therefore requires the presence of nonorthogonal (axial) torsional electric and torsional magnetic fields.

{\it Weyl semimetals with spatially varying dispersion.}---To put the anomaly Eq.~\eqref{eq:axial-torsional-anomaly} in the context of a specific system, we consider a minimal linear model of a Weyl semimetal consisting of two Weyl nodes of opposite chirality, separated in momentum space by a reciprocal vector $2K_i=(0,0,2K)$. 
The Hamiltonian for each chirality reads
\begin{align}
\mathcal{H}_{L/R}&=\frac{i\hbar v}{2}\Big[\bar{\Psi}\big(\underline{e}_{t}^{L/R,i}\pm\sigma^j\underline{e}_j^{L/R,i}\big)\partial_i\Psi \notag \\
&-(\partial_i\bar{\Psi})\big(\underline{e}_t^{L/R,i}\pm\sigma^j\underline{e}_j^{L/R,i)}\big)\Psi\Big],
\label{WeylHamiltonian}
\end{align}
where $\sigma^j$ are the Pauli matrices. 
The term $v\underline{e}_{j}^{L/R,i}$ is a generalized anisotropic Fermi velocity for each chirality, whereas $v\underline{e}_{t}^{L/R,i}$ tilts the Weyl cones~\cite{TSBB15, SB15,XZZ15}. 
The frame field notation naturally accounts for inhomogeneities by allowing the tilt and Fermi velocity to depend on space~\footnote{Notice that the way we are writing the Hamiltonian ensures hermiticity even in the inhomogeneous case.}, in which case both quantities can be seen as a distortion of the geometry of the medium in which the Weyl fermions move~\cite{ZV17,WO17,GYYY17}. 
It is precisely a frame field, as in Eq.~\eqref{WeylHamiltonian} but without chirality dependence, which gives the coupling of Weyl fermions to the background geometry in the standard field theoretical formalism describing Weyl fermions in curved space~\cite{HHKN76,HLP13,PHL14}. 
In our model, in contrast, each chirality is allowed to couple differently to the geometry and there is no spin connection. 
The latter feature means that the inhomogeneous tilt and Fermi velocity are only equivalent to a distortion of space when the spacetime curvature vanishes; although not necessary for our results, all configurations we consider have vanishing curvature.

\begin{figure}
\centering\includegraphics[trim=49cm 30cm 40cm 30cm, clip=true,width=0.5\textwidth]{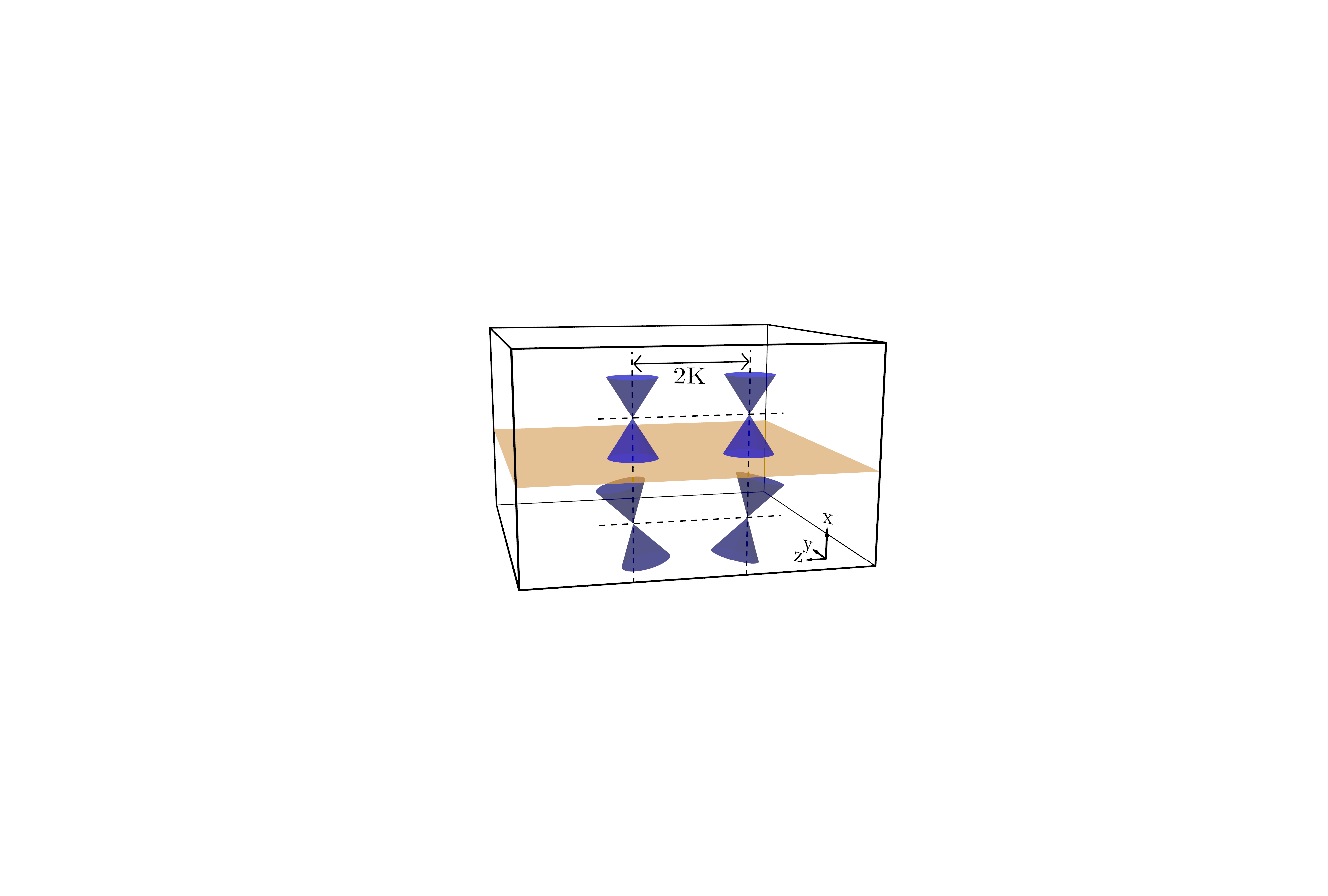}

\caption{Weyl semimetal heterostructure with a tilted interface. Two slabs of Weyl semimetal are stacked along the $x$ axis, both of them with two Weyl nodes separated by identical distances of $2K$ along the $z$ axis in reciprocal space. There is a finite inversion symmetric tilt along the $z$ direction in the lower slab, while the tilt vanishes in the upper one.}
\label{setup}
\end{figure}

We are now in a position to consider a possible realization of axial torsion. 
Take the boundary of the Weyl semimetal~\eqref{WeylHamiltonian}, with a tilt along the $z$ direction and an isotropic and homogeneous Fermi velocity $v$. 
We model a boundary at $x=0$ by the space dependent Weyl node separation vector $K_z=K\Theta(x)$ and tilt $\underline{e}_{t}^{L/R,z}=\mp r\Theta(x)$, with $r$ a constant. 
The gradient in $K_z$ gives an axial magnetic field $B^5_y=\hbar K\delta(x)$ at the surface, the zeroth Landau level of which are the Fermi arcs \cite{GVVI16}, whereas the tilt gradient gives the desired axial torsional magnetic field $\mathcal{B}^{5,t}_y=r\delta(x)$. 
In order to isolate the torsional field, we take an interface between the above defined tilted Weyl semimetal, and the same without the tilt, as represented in Fig.~\ref{setup}. 
The tilt gradient still gives rise to $\mathcal{B}^{t,5}_y$ at the interface, but, crucially, the absence of a gradient in the Weyl node separation means that the axial magnetic field vanishes, and there are no Fermi arcs at the interface. 
Such a stacked Weyl semimetal configuration will be used further down for the activation of the axial-torsional anomaly, and will be important to isolate the torsional contribution from the axial gauge field contribution.
For practical purposes it is enough that the torsional contribution dominates, so it is sufficient that the gradient of tilt is considerably bigger than the gradient of Weyl node separation, relaxing the strict condition of equal Weyl node separation across the interface.

\textit{Strain in Weyl semimetals}.--- In the model Hamiltonian~\eqref{WeylHamiltonian}, the microscopic origin of the inhomogeneous tilt and Fermi velocities was not specified; we now argue that both arise from the application of strain. 
In the continuum limit, displacements of the atoms in a solid are captured by the displacement vector $u^i(x,y,z)$. 
An inhomogeneous displacement vector field generates strain, and nonzero strain indicates that the spatial geometry of the elastic medium has been distorted. 
In fact, the change in the spatial components of the metric is given in terms of the displacement vector~\cite{LL86} as $g_{ij}=\delta_{ij}+2u_{ij}$, with $u_{ij}$ the strain tensor $u_{ij}=1/2(\partial_iu_j+\partial_ju_i)$. 

While such strain-based elasticity theory is quite useful, it is not general enough to model all effects generated by the coupling of spin-orbit coupled materials to geometric deformations; the more fundamental quantity of the frame field $\underline{e}_a^\mu$ is needed in this case~\cite{HLP13}. 
From a lattice point of view, the frame is a set of four vectors residing on each lattice site at any given time, encoding the local bond stretching through their spatial lengths, and the local orbital orientation through their relative angles. 
Importantly, while the metric does not capture local orbital deformations, the frame field does. 
This modification of elasticity theory is related to micropolar or ``Cosserat" elasticity~\cite{E67}. 
To first order in the displacement vector, the frame and coframe fields are given by~\cite{HLP13}  $\underline{e}_a^i=\delta_a^i-\delta_{ak}\,\partial^i u^{k}$ and $e_i^a=\delta_i^a+\delta_{k}^a\,\partial_i u^k$.

The presence, in a Weyl semimetal, of the vector scale $K_i$ that gives the node separation, implies that the above expression for the frame field is not sufficient to encode all the effects of strain.
Symmetry arguments~\cite{AV17} and tight-binding derivations~\cite{CFLV15,CZ16} entail that there are two strain terms in the Hamiltonian that can be constructed by contracting the strain tensor with $K_i$: a pseudoscalar $C\,\hbar vK_ju^{ij}k_i$, with $k_i$ the momentum, and a pseudovector (or axial vector) $A^5_i=\beta\,\hbar K_ju_i^{j}$, with $C$ and $\beta$ model-dependent constants. 
Since $C$ has units of length, we write $C=\gamma a$, where $a$ is the typical lattice spacing and $\gamma$ is a dimensionless model dependent parameter; $\beta$ is a dimensionless constant that in tight-binding calculations~\cite{CFLV15} is equal to the Gr\"uneisen parameter, which is a measure of a crystal's sensitivity to strain, of the given model.
The ``pseudo'' nature of these terms implies they couple with opposite sign to the two chiralities. 
The pseudoscalar term contributes to $\underline{e}_{0}^{i}$ and tilts the Weyl cones, whereas the pseudovector term acts as an axial gauge field, called the elastic gauge field~\cite{VKG10}. 
Hence, the strained system is described by the Hamiltonian~\eqref{WeylHamiltonian}, fixing the Weyl node separation to $2K_i=(0,0,2K)$, with the modified frame field
\begin{equation}
\underline{e}_{a}^{L/R,i}=\delta_a^i-\delta_{ak}\,\partial^i u^{k}\mp \gamma aK\delta_a^t\delta^i_ku^k_z,
\end{equation}
coframe field 
\begin{equation}
e_i^{L/R,a}=\delta_i^a+\delta_{k}^a\,\partial_i u^k\pm\gamma aK\delta_t^a\delta_i^k u^z_k,
\end{equation}
and minimal axial coupling to the elastic gauge field 
\begin{equation}
\hbar\partial_i\rightarrow\hbar\partial_i\pm iA^5_i,\quad A^5_i=\hbar\beta Ku^z_{i}.
\end{equation}

\textit{Realization of the mixed axial-torsional anomaly}.---Having established a possible microscopic origin for the chiral frame fields, we return to the heterostructure of Fig.~\ref{setup}, with constant Weyl node separation but varying tilt. 
One way to achieve this is to stack (here in the $x$-direction) two Weyl semimetals with vanishing tilt but different Weyl node separation (here in the $z$-direction). 
To make the Weyl node separation similar, we apply an uniaxial strain $u^z=\alpha z$ to one side, where $\alpha=\Delta L/L$ measures the elongation of the crystal. 
This strain modifies the Weyl node separation $K\rightarrow K+A_z^5/\hbar=K(1+\beta\alpha)$ and is tuned to make the separation similar in the two samples.
At the same time, the strained sample gets tilted in the $z$ direction, resulting in a tilt gradient across the interface: $\underline{e}_{t}^{L/R,z}=\mp\gamma aKu^z_z=\mp\gamma aK\alpha\,\Theta(x)$. 
Alternatively, the Weyl node separation can be tuned by the application of a magnetic field through the Zeeman term~\cite{CBB17}. 
Although the above procedure can be generalized to Weyl semimetals with multiple Weyl nodes, it may be technically challenging. 
Encouragingly, proposals for minimal time-reversal breaking Weyl semimetals, with a single pair of Weyl nodes, in magnetic Heusler alloys have been put forward~\cite{WVCB16}; these would be ideal for realizing the tilted interface just described.

The tilt gradient through the interface generates an axial torsional magnetic field $\mathcal{B}^{5,t}_y=\gamma aK\alpha\,\delta(x)$, while the axial magnetic field, which is generated by spatial variation in the node separation, vanishes. 
To activate the anomaly, we additionally need a torsional axial electric field. 
This can be achieved by a component of the displacement vector $u^y(z,t)$, which can be realized by driving transverse sound waves through the crystal, resulting in $u^y(z,t)=u_0\sin(k_sz-\omega t)$, with $k_s=\omega/c_s$ the wave number, $\omega$ the frequency, and $c_s$ the sound velocity. 
Such a displacement vector gives rise to $\mathcal{E}^{5,t}_{y}=\gamma aKu_0\omega^2\sin(k_sz-\omega t)/c_s$. 
It also, in fact, gives rise to an axial electric field $E^5_{y}$, but since the axial magnetic field vanishes, the axial gauge field contribution to the anomaly also vanishes. 
The mixed axial-torsional anomaly is therefore the only contribution to the anomaly in this setup.

We solve the torsional anomaly equation for the axial charge density assuming $\partial_iJ_5^i=0$. 
In the presence of intervalley scattering with scattering time $\tau_v$~\cite{PGAPV14}, the anomaly equation takes the form $\partial_tn_5=\vec{\mathcal{E}}_a^{\,5}\cdot\vec{\mathcal{B}}^{\,5,a}/2\pi^2l^2_5-n_5/\tau_v$, where $n_5=J^0_5/e$ is the axial number density. 
Inserting the explicit form of the torsional fields and solving for the density in the limit where the phonon frequency is much larger than the intervalley scattering rate $\omega \tau_v\gg1$, we get, at long times $t\gg\tau_v$,
\begin{equation}
n_5=-\frac{\gamma^2K^2\alpha\,u_0\,\omega}{2\pi^2c_s}\cos(k_sz-\omega t)\,\delta(x).
\label{eq:torsionalanomaly2}
\end{equation}
In arriving at Eq.~\eqref{eq:torsionalanomaly2} we have taken the cut-off length scale $l_5=a$ equal to the lattice spacing, the physical cut-off length scale of the crystal.
Notice that the (co)frame field vanishes outside the material and therefore its total flux through the sample must vanish $\int dxdz\,\mathcal{B}^{5,t}_y=0$~\cite{HQ13}. 
Consequently, there must be a contribution to the anomaly localized at the lower surface [Fig.~\ref{setup}], where the tilt gradient has a sign opposite to that at the interface, such that the total (spatially integrated) axial number density is conserved \footnote{The lower (and upper) surface presents an axial gauge field contribution to the anomaly as well (note that the phonons generate an axial electric field). Only at the interface is the torsional contribution isolated.}.

\begin{figure}
\centering\includegraphics[trim=45cm 31cm 38cm 27cm, clip=true,width=0.5\textwidth]{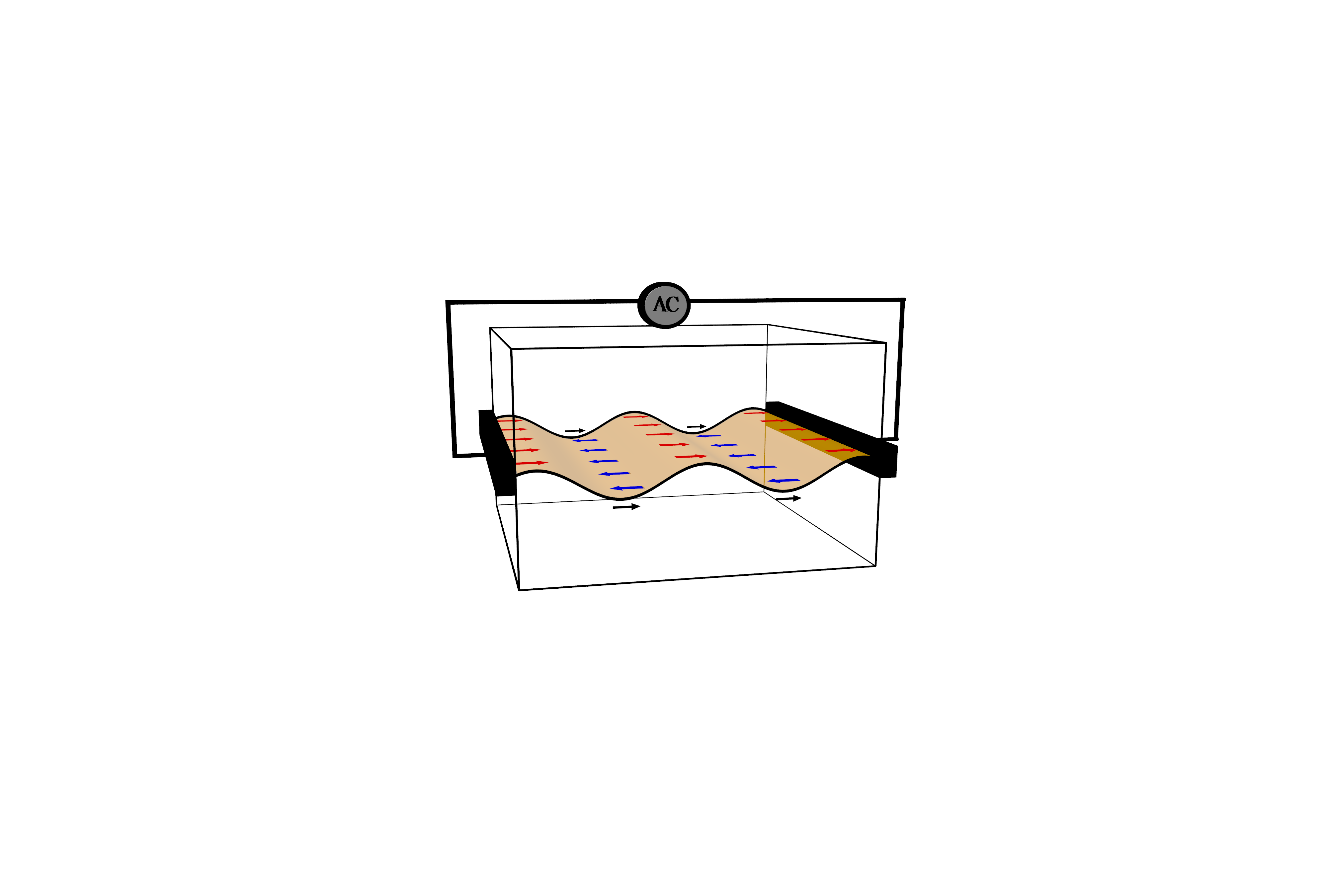}

\caption{The mixed axial-torsional anomaly is activated by driving transverse phonons through the Weyl semimetal heterostructure represented in Fig. \ref{setup}. The subsequent application of a magnetic field in the $z$ direction (horizontal axis in the picture) yields the above depicted charge density wave (CDW) at the tilted interface. The sinusoidal pattern represents the magnitude of the charge at each point in the interface. The red and blue arrows represent the direction of the current at the points where the CDW amplitude has its maxima and minima, respectively, while the black arrows point towards the direction of propagation of the CDW. Two leads are placed at each side of the interface, such that the propagating CDW generates an alternating current through the circuit.}
\label{CDW}
\end{figure}

\textit{Experimental detection}.---To experimentally detect the spacetime oscillating axial charge~\eqref{eq:torsionalanomaly2}, we make use of the chiral magnetic effect~\cite{FKW08,LKV16}.
Due to this effect, a magnetic field in the $z$ direction~\footnote{Notice that if the Zeeman coupling is of different strength in the two materials, then the magnetic field would induce different Weyl node separations in the two samples. The strain would need to be adjusted to correct for this effect.} induces a current $J^{z}_\mathrm{CME}=e^2\mu_5B/2\pi^2\hbar^2$ parallel to the applied magnetic field, where $\mu_5=(\mu_L-\mu_R)/2$ is the axial chemical potential. 
In the weak field limit $\hbar eB\ll\mu_5^2/v^2$, $n_5$ is related to $\mu_5$ according to $3\pi^2\hbar^3v^3n_5=\mu_5^3+\mu_5(\pi^2\kappa_B^2T^2+\mu^2)$, where $T$ and $\mu$ are the temperature and chemical potential~\cite{FKW08}. 
In the realistic limit $\mu,\kappa_BT\gg\mu_5$ we can drop the $\mu_5^3$ term. 
Crucially, the chiral magnetic current $J^{z}_\mathrm{CME}$, being proportional to the axial chemical potential, is only nonzero if the axial anomaly is activated. 

The two-dimensional chiral magnetic current density at the interface between the two Weyl semimetals, using Eq.~\eqref{eq:torsionalanomaly2}, now reads
\begin{equation}
J=\int dx\,J^z_\mathrm{CME}=-\frac{3\hbar^3v^3\gamma^2K^2\alpha\,u_0\,\omega B}{2c_s(\pi^2\kappa_B^2T^2+\mu^2)}\cos(k_sz-\omega t).
\end{equation}
Due to current conservation $\partial_\mu J^\mu=0$, this generates a propagating charge density wave with charge density $\rho=J^0=J/c_s$. 
The mixed axial-torsional anomaly can then be tested by measuring the AC current $J$ flowing through current leads placed at opposite sides of the interface [Fig. \ref{CDW}]. 
For typical values $v=10^6$ m/s, $K=10^9$ m$^{-1}$, $\gamma=1$, $\mu=10$ meV, for strain of $3\%$, phonon amplitude $u_0=a/10$, with lattice constant $a=5$ \r{A}, sound speed $c_s=2\times10^{3}$ m/s, and driving frequency $\omega=1$ THz, we estimate the amplitude of the high frequency AC current at room temperature to be $J=0.9\,B$ A/m, with $B$ in Tesla. 
For a magnetic field of $B=10$ mT, which fulfills the weak field condition, we thus obtain $J=9$ mA/m. 
Assuming a sample given by a ribbon of 5 $\mu$m width, the amplitude of the total current that passes through the leads is then $J=45$ nA, which is well within the range of existing experimental probes. 
The linear dependence of the current amplitude with the magnetic field serves as well to distinguish it from noise.

{\it Discussion.---} In this work we have demonstrated that a hitherto overlooked anomaly, the mixed axial-torsional anomaly with axial torsion, is naturally realised in a condensed matter setting. 
In particular, we demonstrated that this anomaly should be detectable, within the current experimental capabilities, by driving transverse sound waves through tilted Weyl semimetal interfaces, and measuring the induced alternating currents in the presence of an external magnetic field. For driving phonon frequencies in the THz regime, and a small value of the magnetic field of $10$ mT, we predict a current amplitude of around $40$ nA.

Our treatment of the axial-torsional anomaly with axial torsion opens the path for a more in-depth study of anomaly induced torsional responses in Weyl systems. 
On the experimental and phenomenological side, other implementations of inhomogeneities, such as magnetization in magnetic Weyl semimetals, which could give rise to nonvanishing torsion are worth exploring. 
Another appealing direction is to extend the methodology presented here to study the realization of a mixed axial-gravitational anomaly, to come up with a realization of axial curvature, and study how it modifies the gravitational contributions to the anomaly.
All these effects could in principle be engineered in Weyl semimetals.

{\it Acknowledgments.---} This work was supported by the ERC Starting Grant No. 679722 and the Knut and Alice Wallenberg Foundation 2013-0093. Y.K. and E.J.B. are supported by the Swedish research council (VR) and the Wallenberg Academy Fellows program of the Knut and Alice Wallenberg Foundation.

\appendix
\section{Derivation of the mixed axial-torsional anomaly with axial torsion}
\label{appendix}

Here we derive the expression for the mixed axial-torsional anomaly in the presence of axial torsion fields, and for vanishing curvature. But before going into the actual derivation, we need some definitions. Let us introduce the affine connection, $\tensor{\Gamma}{_\mu_\nu^\rho}$, and write it down as $\tensor{\Gamma}{_\mu_\nu^\rho}=\tensor{\bar{\Gamma}}{_\mu_\nu^\rho}+\tensor{C}{_\mu_\nu^\rho}$. The first contribution is the Levi-Civita connection, which for vanishing curvature can be chosen to vanish everywhere. The second contribution is the contorsion tensor, related to the torsion tensor as $2\tensor{C}{_\mu_\nu^\rho}=\tensor{T}{_\mu_\nu^\rho}+\tensor{T}{_\nu^\rho_\mu}+\tensor{T}{^\rho_\mu_\nu}$. See that $C$ is antisymmetric in its last two indexes. We can now define, in flat space, the covariant derivative acting on contravariant vectors $\nabla_\mu V^\nu=\partial_\mu V^\nu+\tensor{C}{_\gamma_\mu^\nu}V^\gamma$ and second rank tensors $\nabla_\mu V^{\nu\rho}=\partial_\mu V^{\nu\rho}+\tensor{C}{_\gamma_\mu^\nu}V^{\gamma\rho}+\tensor{C}{_\gamma_\mu^\rho}V^{\nu\gamma}$. It follows that the covariant divergence of a vector is equal to the ordinary divergence, $\nabla_\mu V^\mu=\partial_\mu V^\mu+\tensor{C}{_\gamma_\mu^\mu}V^\gamma=\partial_\mu V^\mu$, whereas the covariant divergence of a tensor is $\nabla_\mu V^{\nu\mu}=\partial_\mu V^{\nu\mu}+\tensor{C}{_\gamma_\mu^\nu}V^{\gamma\mu}$.

Now we are ready to compute the anomaly in the presence of axial torsion fields. We start from the known expression of the axial-torsional anomaly in the absence of axial fields ($e^{a(L)}_\mu=e^{a(R)}_\mu$) \cite{CZ97,PHL14}
\begin{equation}
\partial_\mu J^\mu_5=\frac{e}{16\pi^2l^2}\epsilon^{\mu\nu\rho\lambda}T^a_{\mu\nu}T^b_{\rho\lambda}\eta_{ab},
\label{eqsupp:anomaly1}
\end{equation}
with $l$ an UV cut-off and $T_{\mu\nu}^a=\partial_\mu e^a_{\nu}-\partial_\nu e^a_\mu$ the torsion tensor in the absence of background curvature. In terms of the individual anomalies for left- and right-handed fermions, Eq. \eqref{eqsupp:anomaly1} reads
\begin{equation}
\partial_\mu J^\mu_{L,R}=\pm\frac{e}{32\pi^2l^2}\epsilon^{\mu\nu\rho\lambda}T^{a(L,R)}_{\mu\nu}T^{b(L,R)}_{\rho\lambda}\eta_{ab}.
\label{eqsupp:anomaly2}
\end{equation}
with $T^{a(L,R)}_{\mu\nu}=\partial_\mu e^{a(L,R)}_{\nu}-\partial_\nu e^{a(L,R)}_\mu$. Here the anomaly lies entirely in the U(1) sector. It breaks gauge invariance of chiral fermions, but respects diffeomorphism symmetry. This means that the energy-momentum tensor is conserved, $\nabla_\mu\mathcal{T}^{\nu\mu}=0$. Notice that the covariant divergence of $\mathcal{T}^{\nu\mu}$ can be written as $\nabla_\mu (\underline{e}^\nu_a\mathcal{T}^{a\mu})=\nabla_\mu\mathcal{T}^{a\mu}$, where we have used the fact that $\nabla_\mu\underline{e}^\nu_a=0$, which is usually called the tetrad postulate. The energy-momentum tensor, $\mathcal{T}^{a\mu}$, is defined as the functional derivative of the action with respect to the coframe field
\begin{equation}
\mathcal{T}^{a\mu}=\frac{1}{\det e}\frac{\delta S}{\delta e_\mu^b}\eta^{ab}.
\end{equation}

Next we allow for $e^{a(L)}_\mu\neq e^{a(R)}_\mu$ and define: $e^{a}_\mu=(e^{a(L)}_\mu+e^{a(R)}_\mu)/2$, $e^{5,a}_\mu=(e^{a(L)}_\mu-e^{a(R)}_\mu)/2$, $T^{a}_{\mu\nu}=T^{a(L)}_{\mu\nu}/2+T^{a(R)}_{\mu\nu}/2=\partial_\mu e^a_{\nu}-\partial_\nu e^a_\mu$, $T^{5,a}_{\mu\nu}=T^{a(L)}_{\mu\nu}/2-T^{a(R)}_{\mu\nu}/2=\partial_\mu e^{5,a}_{\nu}-\partial_\nu e^{5,a}_\mu$. Now, in terms of vector and axial currents Eq. \eqref{eqsupp:anomaly2} becomes 
\begin{equation}
\partial_\mu J^\mu=\frac{e}{8\pi^2l^2}\epsilon^{\mu\nu\rho\lambda}T^a_{\mu\nu}T^{5,b}_{\rho\lambda}\eta_{ab},
\label{eqsupp:anomaly3}
\end{equation}
\begin{equation}
\partial_\mu J^\mu_5=\frac{e}{16\pi^2l^2}\epsilon^{\mu\nu\rho\lambda}\Big(T^a_{\mu\nu}T^b_{\rho\lambda}+T^{5,a}_{\mu\nu}T^{5,b}_{\rho\lambda}\Big)\eta_{ab}.
\label{eqsupp:anomaly4}
\end{equation}
We see that because of a finite $T^{5}$, the divergence of the axial current acquires an additional axial field contribution, and what is more worrying, the vector current is no longer conserved. This is telling us that, in the presence of $T^{5}$,  our left-right symmetric regularization is not physical, and that we must choose an alternative, gauge invariant, regularization such that the vector current is conserved. This can be done by adding (non-gauge invariant) local counterterms to the action, called Bardeen counterterms \cite{B69}, with the additional caveat that the diffeomorphism symmetry has to be respected in the process. This is at the end, the condition $\partial_\mu J^\mu=\nabla_\mu\mathcal{T}^{a\mu}=0$ must be fulfilled.

Following the above argumentation, there is a Bardeen counterterm that cancels the vector current non-conservation of Eq. \eqref{eqsupp:anomaly3} while still respecting the conservation of the energy-momentum tensor. This term is
\begin{equation}
S_{c.t.}=-\frac{e}{4\pi^2l^2}\int d^4x\,\epsilon^{\mu\nu\rho\lambda}A_\mu e^{5,a}_\nu T^{b}_{\rho\lambda}\eta_{ab}.
\label{eqsupp:anomaly5}
\end{equation}
If we compute the divergence of the current and energy-momentum tensor arising from $S_{c.t.}$ we get
\begin{equation}
\partial_\mu J^\mu=\partial_\mu\frac{\delta S_{c.t.}}{\delta A_\mu}=-\frac{e}{8\pi^2l^2}\epsilon^{\mu\nu\rho\lambda}T^a_{\mu\nu}T^{5,b}_{\rho\lambda}\eta_{ab},
\end{equation}
\begin{equation}
\nabla_\mu T^{a\mu}=\nabla_\mu\frac{1}{\det e}\frac{\delta S_{c.t.}}{\delta e_\mu^b}\eta^{ab}=0.
\end{equation}
Such a Bardeen counterterm switches the anomaly from the vector current to the axial energy-momentum tensor, which is no longer conserved. The consequences of a non-conservation of the axial energy-momentum tensor will be discussed elsewhere. After adding the countribution in Eq. \eqref{eqsupp:anomaly5} we get
\begin{equation}
\partial_\mu J^\mu=0,
\end{equation}
\begin{equation}
\partial_\mu J^\mu_5=\frac{e}{16\pi^2l^2}\epsilon^{\mu\nu\rho\lambda}\Big(T^a_{\mu\nu}T^b_{\rho\lambda}+T^{5,a}_{\mu\nu}T^{5,b}_{\rho\lambda}\Big)\eta_{ab}.
\end{equation}

Beyond the term defined in Eq. \eqref{eqsupp:anomaly5}, mandatory to recover gauge invariance, there is another Bardeen counterterm that can be defined while respecting both gauge and diffeomorphism invariance. It mixes axial gauge and frame fields, and reads
\begin{equation}
\frac{e\,C_1}{8\pi^2l^2}\int d^4x\epsilon^{\mu\nu\rho\lambda}A^5_\mu e^{5,a}_\nu T^{5,b}_{\rho\lambda}\eta_{ab}
\end{equation}
where $C_1$ is an arbitrary, regularization dependent constant. This term contributes to the divergence of the axial current as
\begin{equation}
\partial_\mu J^\mu_5=\frac{e}{16\pi^2l^2}\epsilon^{\mu\nu\rho\lambda}\Big(T^a_{\mu\nu}T^b_{\rho\lambda}+(1+C_1)T^{5,a}_{\mu\nu}T^{5,b}_{\rho\lambda}\Big)\eta_{ab}.
\end{equation}
Absorbing the constant $C_1$ into a new length scale $l_5^2=l^2/C_1$ we get the mixed axial-torsional anomaly as presented in the main text
\begin{equation}
\partial_\mu J^\mu_5=\frac{e}{16\pi^2l^2}\epsilon^{\mu\nu\rho\lambda}\Big(T^a_{\mu\nu}T^b_{\rho\lambda}+\frac{l^2}{l_5^2}T^{5,a}_{\mu\nu}T^{5,b}_{\rho\lambda}\Big)\eta_{ab}.
\end{equation}

\bibliography{torsionLib}

\end{document}